\documentclass[a4paper,fleqn]{cas-dc}

\usepackage[]{natbib}
\usepackage{amsmath}
\usepackage[amsmath,thref]{ntheorem}
\usepackage{listings}
\usepackage{nomencl}
\makenomenclature
\usepackage{geometry}
\usepackage{bm}
\usepackage[section]{placeins}
\usepackage{multirow}
\usepackage{tikz}

\usetikzlibrary{tikzmark}

\definecolor{vert}{rgb}{0,0.5,0}

\usepackage{etoolbox}
\renewcommand\nomgroup[1]{%
  \item[\bfseries
  \ifstrequal{#1}{V}{Variables}{%
  \ifstrequal{#1}{P}{Parameters}{%
  \ifstrequal{#1}{G}{Greek letters}{%
  \ifstrequal{#1}{S}{Subscripts}{%
  \ifstrequal{#1}{A}{Acronyms}{}}}}}%
]}
\usepackage{tabularx,booktabs}
\usepackage{multicol}
\usepackage[switch]{lineno} 
  
\allowdisplaybreaks

\def\tsc#1{\csdef{#1}{\textsc{\lowercase{#1}}\xspace}}
\tsc{WGM}
\tsc{QE}
\tsc{EP}
\tsc{PMS}
\tsc{BEC}
\tsc{DE}
\begin{document}
 
\let\WriteBookmarks\relax
\def\floatpagepagefraction{1}
\def\textpagefraction{.001}

\shorttitle{Analytical expressions for the MILS model}
\shortauthors{Pasquier P. and Lamarche L.}

\title [mode = title]{Analytic expressions for the moving infinite line source model}                

\author[1]{Philippe Pasquier}
\cortext[cor1]{Corresponding author}
\ead{philippe.pasquier@polymtl.ca}
\credit{Conceptualization, Methodology,  Validation,  Software, Investigation, Writing - Original Draft, Writing - Review \& Editing, Visualization}

\address[1]{Department of Civil, Geological and Mining Engineering, Polytechnique Montr\'{e}al, P.O. Box 6079 Centre-Ville, Montr\'{e}al, Qu\'{e}bec, Canada H3C 3A7}

\author[2]{Louis Lamarche} 
\credit{Validation, Investigation, Writing - Review \& Editing}

\address[2]{Department of Mechanical Engineering, ÉTS, 1100, Notre-Dame W., Montr\'{e}al, Qu\'{e}bec, Canada H3C 1K3}

\begin{abstract}
Groundwater flow can have a significant impact on the thermal response of ground heat exchangers. The moving infinite line source model is thus widely used in practice as it considers both conductive and advective heat transfert processes. Solution of this model involves a relatively heavy numerical quadrature. Contrarily to the infinite line source model, there is currently no known first-order approximation that could be useful for many practical applications. In this paper, known analytical expressions of the Hantush well function and generalized incomplete gamma function are first revisited.  A clear link between these functions and the moving infinite line source model is then established.  Then, two new exact and integral-free analytical expressions are proposed, along with two new first-order approximations. The new analytical expressions proposed take the form of convergent power series involving no recursive evaluations. It is shown that relative errors less than 1\% can be obtained with only a few summands. The convergence properties of the series, their accuracy and the validity domain of the first-order approximations are also presented and discussed. 
\end{abstract}

% \includegraphics{figs/grabs.pdf}

% \begin{highlights}
% \item Analytic expressions for the moving infinite line source model are presented
% \item The power series are convergent and accuracy of 1\% is achieved with a few terms
% \item First-order approximations and their validity range are also provided
% \end{highlights}


\maketitle

 \section{Introduction}
By promoting advective heat transfer, groundwater flow can have a significant impact on ground-source heat pump and borehole thermal energy storage systems. Indeed, research illustrating how groundwater flow can influence mean fluid temperatures \citep{chiasson_preliminary_2000-1}, thermal performances \citep{fan_study_2007,zanchini_long-term_2012,nguyen_borehole_2017}, geometrical layouts \citep{choi_numerical_2013} or operation costs \citep{capozza_investigations_2013,samson_influence_2018} of these systems are numerous. The result of thermal response tests being also impacted by the advection of groundwater \citep{signorelli_numerical_2007,chiasson_new_2011,angelotti_energy_2014}, much work has been devoted to parameters identification with various interpretation models \citep{raymond_numerical_2011, wagner_analytical_2013, antelmi_thermal_2020,magraner_thermal_2021}. 

\begin{figure}%   trim={<left> <lower> <right> <upper>}
 \centering
  \includegraphics[width=0.5\textwidth,trim={3cm 5cm 2.25cm 5cm},clip]{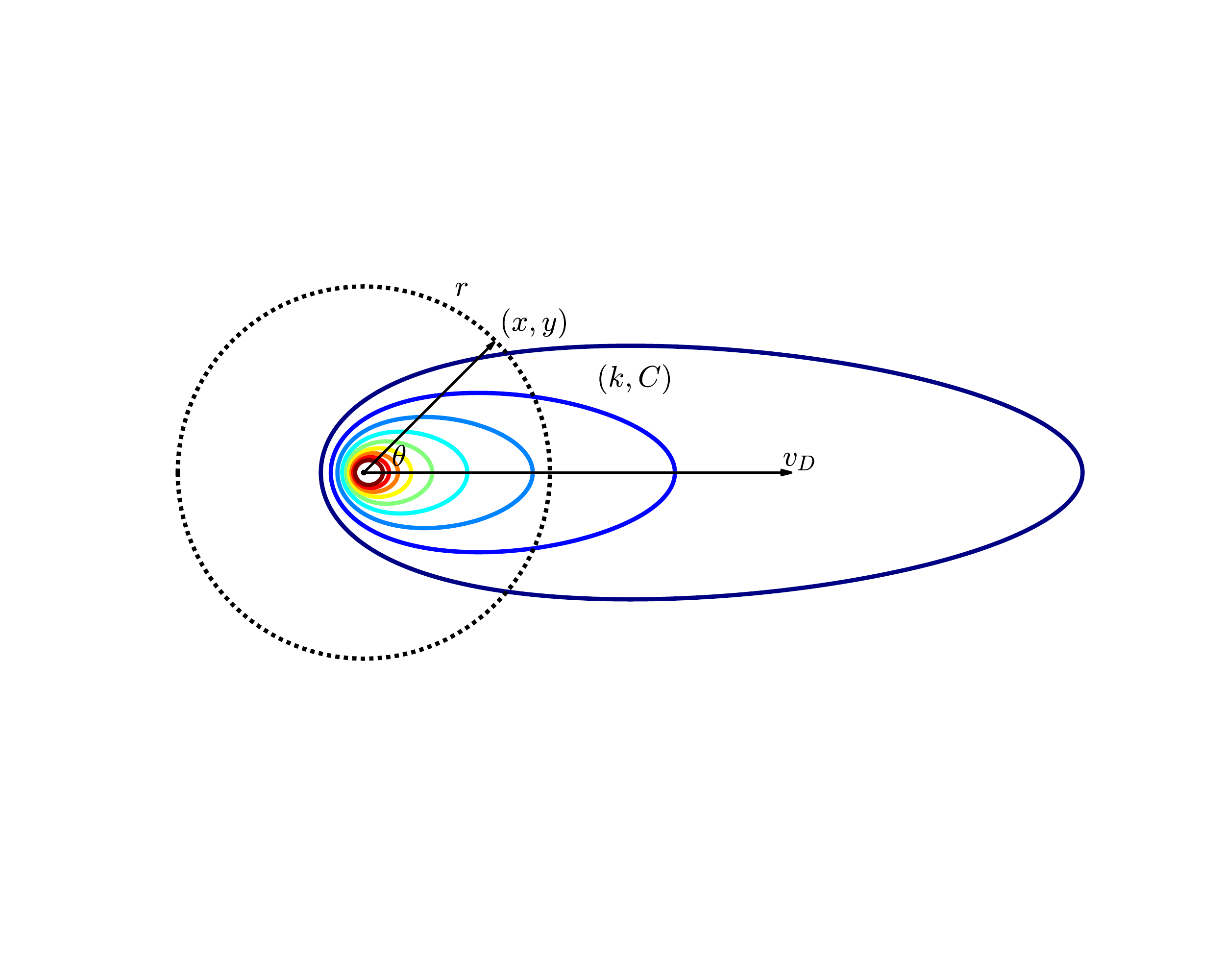}
 \caption{Illustration of the thermal plume caused by groundwater advection downstream of an infinite line heat source. Warm temperatures correspond to red colors.}
 \label{FIG:1}
\end{figure}

It results that if a significant groundwater flow is present in the aquifer, the design and analysis of ground heat exchangers (GHE) must then be accomplished with simulation models considering both conduction and advection of groundwater. Such a model, known as the moving infinite line source (MILS) model, was first proposed by \citet[p. 267]{carslaw_conduction_1959} for steady-state. The first known extension to the time-dependent case is due to \citet{zubair_temperature_1996} who also proposed some other interesting solutions and established a clear link with the generalized incomplete  gamma  function $\Gamma(0,x;b)$. The radial evolution of the temperature around the line source was studied by \citet{zeng_moving_1997} who observed greater temperature variations downstream of the heat source. However, the first applications of the MILS to GHEs are due almost simultaneously to \cite{sutton_ground_2003} and \cite{diao_heat_2004} for design and analysis of ground-source heat pump systems. Nowadays, the general form of the MILS model is:
\begin{equation}\label{eq:1}
\Delta T=\frac{q}{4 \pi k} e^{\frac{r \cos{\theta} \, v_{T} }{2 \alpha }  } \int \limits_{r^2 /4\alpha t}^{\infty} \frac{1}{\psi} e^{-\psi- \frac{1}{\psi}\left (\frac{r \, v_T }{4\alpha } \right)^2 }  d\psi
\end{equation}
with the effective heat transport velocity $v_T$ given by
\begin{equation}\label{eq:2}
v_T=\frac{C_w}{C}v_D
\end{equation}
\noindent Solution of Eq. \ref{eq:1} is illustrated in Fig. \ref{FIG:1} for various coordinates. One can see the thermal plume, which extends mainly in the flow direction.

The mean temperature being useful for design purposes, \cite{sutton_ground_2003} and \cite{diao_heat_2004} used the identity $\int_{0}^{2\pi}e^{x \cos{\theta}}d\theta=2\pi I_0(x)$ to predict it on a circle of radius $r$ centered on the line source. The resulting mean temperature is then given by
\begin{equation}\label{eq:3}
\Delta \overline{T}=\frac{q I_0(r\, v_{T}  / 2 \alpha )}{4 \pi k} \int \limits_{r^2 /4\alpha t}^{\infty} \frac{1}{\psi} e^{-\psi- \frac{1}{\psi}\left (\frac{r \, v_T }{4\alpha } \right)^2 }  d\psi
\end{equation}

\noindent where $I_0$ is the modified bessel function of the first kind and of order 0.

Equation \ref{eq:3} gives the mean temperature change around an infinite line-heat source embedded in a homogeneous, isotropic and infinite media where no \textcolor{black}{hydromechanic} dispersion is active. It is important to note that the thermal parameters $k$, $C$ and $\alpha$ represent equivalent values, that is weighted averages of the thermal parameters of the solid and fluid \citep{nield_convection_2006}.  Thus, $\Delta \overline{T}$ is obtained under the assumption of local thermal equilibrium, which implies that the temperature of the fluid and mineral phases are equal locally. A special attention should also be paid to the fact that $v_T$ is based on the groundwater flux or Darcy velocity $v_D$ \citep{bear_hydraulics_1979}, and not on the \textcolor{black}{ actual pore water velocity}. 

In order to widen its application field, several improvements have been brought to the MILS model. \cite{molina-giraldo_moving_2011} extended the model to simulate line sources of finite length. \cite{metzger_optimal_2004} added the effect of \textcolor{black}{hydromechanic} dispersion in the porous media, a development used later by \cite{molina-giraldo_evaluating_2011} to simulate GHEs response. \cite{rivera_finite_2016} added a Cauchy boundary condition along the ground surface to better reproduce surface conditions. \cite{hu_improved_2017} and \cite{erol_multilayer_2018} proposed models to take into account the multilayer case, thus lifting the assumption of a homogeneous media.  \textcolor{black}{Finally, \cite{van_de_ven_enhanced_2021} recently developed a correction function to the MILS model to take into account the impervious grout put in place in GHEs.}

So far, the MILS model involves an integral for which no closed-form expression is known in the geoexchange community. Thus, solving this integral requires a numerical quadrature that prevents using widely available spreadsheets and limits the efficiency of optimization methods for GHE design or thermal response test analysis. Lack of simple explicit analytical expressions also limits the physical interpretation of the MILS model. However, several analytical expressions, that went unnoticed to the geoexchange community, can ease and accelerate the use of the MILS model. Indeed, in the field of hydrogeology the integral of Eq. \ref{eq:3} is known as the Hantush well function  $W(u,r/\beta)$ \citep{hantush_non-steady_1955}, a well-known integral used to describe the hydraulic head within a leaky confined aquifer. The first analytical solutions to this integral were already given in the seminal paper of \cite{hantush_non-steady_1955} and took the form of a combination of power series and special functions. Since, more practical expressions in the form of analytic expressions, asymptotic expansions and approximations have been proposed. \cite{veling_hantush_2010} realized a comprehensive technical review of the Hantush integral and its various solutions. The expressions proposed so far vary in terms of convergence properties and compactness levels, but the two compact solutions proposed by \cite{hunt_calculation_1977} involve the generalized incomplete gamma function $\Gamma(0,x;b)$ and merits to be known.

Apart from the noticeable work of \cite{capozza_investigations_2013} who exploited the properties of $\Gamma(0,x;b)$ to foster the evaluation of the temperature field in a GHE, the relationship with $\Gamma (0, x; b) $ has gone quite unnoticed by the geoexchange community. Also, contrarily to the infinite line source model, there is currently no known first-order approximation of the MILS that could be useful in practice. In this paper, we first aim to revisit and make a clear link between the MILS model, the Hantush well function and the generalized incomplete gamma function. Then, based on Hunt's solutions, two new exact and integral-free analytical expressions to the MILS model are proposed, along with two useful first-order approximations of the MILS model

%Mention an error in Molina integral bound.

%%%%%%%%%%%%%%%%%%%%%%%%%%%%%%%%%%%%%%%%%%%%%%%%%%%%%%%
\section{Hantush Well function and Hunt's solutions}
In hydrogeology, the integral of Eq. \ref{eq:3} is known as the Hantush well function  \citep{hantush_non-steady_1955} and is used to predict the hydraulic head within leaky confined aquifers. Following the work of \cite{simon_numerical_2021}, doing the change of variable 
\begin{equation}
u=\frac{r^2}{4\alpha t}
\end{equation}
\noindent and
\begin{equation} 
\beta=\frac{2\alpha } {v_T}
\end{equation}

\noindent allows to reconcile the MILS model and the usual form of the Hantush well function given by:
\begin{equation}\label{eq:6}
W(u,r/\beta)= \int \limits_{u}^\infty \psi^{-1} e^{\left (-\psi- \frac{r^2}{4\beta^2\psi} \right)  }d\psi
\end{equation}

\noindent Now, by making the change of variable:
\begin{equation} 
 b=\frac{1}{4}\left (\frac{r  }{\beta } \right)^2=\frac{1}{4}\left (\frac{r \, v_T }{2 \alpha } \right)^2 
\end{equation}
\noindent the parallel between $W$ and the generalized incomplete gamma function described by \citet{chaudhry_generalized_1994} becomes clear. The latter reads 
\begin{equation} \label{eq:x}
 \Gamma(a,u;b)   = \int \limits_{u}^\infty \psi^{a-1} e^{\left (-\psi- \frac{b}{\psi} \right)  }d\psi  
 \end{equation}
 \noindent and it is clear that the Hantush well function $W$ corresponds to the case $a=0$. Note that for $b=0$, $\Gamma(a,u;b)$ reduces to the incomplete gamma function while the case $a=b=0$ corresponds to the exponential integral function $E_1$.
 
 Two useful solutions to the Hantush integral were proposed by \cite{hunt_calculation_1977} a few decades ago. Using the previous definition of $b$ and expressing the factor $e^{-b/\psi}$ in Eq. \ref{eq:6} by his power series, one obtains:
\begin{equation} \label{eq:8}
\begin{aligned}[b]
%W(u,b) & = \int \limits_{u}^\infty \psi^{-1} e^{\left (-\psi- \frac{b}{\psi} \right)  }d\psi \\ 
W(u,b) & = \sum _{n=0}^{\infty} \frac{(-b)^n}{n!} \int  \limits_{u}^{\infty} \psi^{-n-1} e^{-\psi} d\psi\\
& = \sum _{n=0}^{\infty} \frac{(-b)^n}{n!} u^{-n}\int  \limits_{1}^{\infty} {\psi'}^{-n-1} e^{-u\psi'} d\psi'\\
& = \sum _{n=0}^{\infty} \frac{(-b/u)^n}{n!} E_{n+1}(u), \,\,\, u>0
\end{aligned}
\end{equation}

\noindent where $E_n$ is the generalized exponential integral of order $n$. \cite{hunt_calculation_1977} mentioned initially that the power series is absolutely convergent for $0\leq b/u <\infty$. This convergence range was however reduced to $u\geq \sqrt{b}$ by \cite{veling_hantush_2010}.  

A second expression provided by \citet{hunt_calculation_1977} is obtained by substituting $\psi'=b/\psi$ in Eq. \ref{eq:6}. One can then derive 
\begin{equation} \label{eq:9}
\begin{aligned}[b]
 W(u,b) & = \int \limits_{0}^{b/u} {\psi'}^{-1} e^{\left (-\psi'- \frac{b}{\psi'} \right)  }d\psi'  \\
& =2K_0\left(2\sqrt{b}\right) - \int \limits_{b/u}^{\infty} {\psi'}^{-1} e^{\left (-\psi'- \frac{b}{\psi'} \right)  } d\psi'\\
& =2K_0\left(2\sqrt{b}\right) - \sum _{n=0}^{\infty} \frac{(-u)^n}{n!} E_{n+1}(b/u), \, u\geq 0
\end{aligned}
\end{equation}
\noindent where $K_0$ is the modified bessel function of the second kind and of order 0. \cite{hunt_calculation_1977} mentioned that the power series is absolutely convergent for $0\leq u < \infty$. Equation \ref{eq:9} was however deemed appropriate by \cite{veling_hantush_2010} for $0< u<\sqrt{b}$. 

Implementation of formulas \ref{eq:8} and \ref{eq:9} is rather easy since the generalized exponential integral function $E_n(x)$ can be obtained recursively with $E_{n+1}(x)=(e^{-x}-x E_n(x))/n$. Since $E_1(x)$ corresponds to the exponential integral function, an evaluation of $E_1$ is needed for $n=0$ and the higher-order terms of $E_n(x)$ are then obtained recursively. For most practical applications, $u$ is not a scalar but a vector whose entries correspond to the various evaluation times. To benefit from the parallelism of modern computers, one has then advantage to use formulas \ref{eq:8} and \ref{eq:9} under a vectorized form. 
 
\section{New analytical expressions}

The recursive evaluation of $E_n$ in Eqs. \ref{eq:8} and \ref{eq:9} does not facilitate their use in spreadsheets and efficiency gains are probably possible by reducing the number of operations achieved on vectors. In this section, two new power series are presented. To ease manipulation and identification of these series, the change of variable  $\tau=4\alpha t/r^2$ , that is four times the Fourier number, has been made. This leads to:
\begin{equation} \label{eq:x}
 \tau=\frac{1}{u}=\frac{4\alpha t}{r^2}
\end{equation}

\textcolor{black}{ 
By setting the characteristic length equal to the radial distance $r$, one can easily establish a link with the Fourier number ($Fo=\alpha t/r^2$) and the thermal Péclet number ($P\acute{e}=rv_T/\alpha$). Indeed, $\tau$ is related to the Fourier number by $\tau=4Fo$ and expresses the conduction rate to the rate of heat storage. Similarly, the parameter $b$ is related to the Péclet number by $b=(P\acute{e}/4)^2$ and expresses the advective heat transfer rate to the conductive heat transfer rate. Although the dimensionless thermal numbers $Fo$ and $P\acute{e}$ are widely used, the use of variables $\tau$ and $b$ leads to more compact analytical expressions. For conciseness, the dimensionless numbers $Fo$ and $P\acute{e}$ will therefore not be used in this Section.}

\subsection{Analytical expression for $\boldsymbol{ 0 < \tau \leq 1/b}$}
Using the dimensionless variables $b$ and $\tau$ in Eq. \ref{eq:8} and expanding the sum gives:
\begin{equation} \label{eq:x}
W(\tau,b)=\sum _{n=0}^{\infty} \frac{(-b\tau)^n}{n!} E_{n+1}(1/\tau) 
\end{equation}
\noindent and 
\begin{equation} \label{eq:x}
\begin{aligned}[b]
 W(\tau,b)= & +\frac{E_1(1/\tau)}{1}-(b\tau)\frac{E_2(1/\tau)}{1}\\ 
 & +(b\tau)^2\frac{E_3(1/\tau)}{2} -(b\tau)^3\frac{E_4(1/\tau)}{6}\\
 & +(b\tau)^4\frac{E_5(1/\tau)}{24} -...
\end{aligned}
\end{equation}

\noindent The order of the generalized exponential integral $E_n$ can be lowered using the recurrence relation \textcolor{black}{ $(e^{-x}- nE_{n+1}(x))/x = E_n(x)$}. Thus, using this relation iteratively to get rid of the generalized exponential integral of order $n>1$ leads to the following continued fraction: 
\begin{small}
\begin{align*}
W(\tau,b) =   +\frac{{E_1}\left(1/\tau\right)}{1} \\
  -(b\,\tau) \,\frac{{\left({\mathrm{e}}^{-\frac{1}{\tau }} -\frac{E_1\left(1/\tau \right)}{\tau }\right)}}{1} \\
  +(b \,\tau)^2 
\frac{{\left(\frac{{\mathrm{e}}^{-\frac{1}{\tau }} }{2}-\frac{{\mathrm{e}}^{-\frac{1}{\tau }} -\frac{E_1\left(1/\tau\right)}{\tau }}{2\,\tau }\right)}}{2} \\
  -(b \,\tau)^3 \, \frac{{\left(\frac{{\mathrm{e}}^{-\frac{1}{\tau }} }{3}-\frac{\frac{{\mathrm{e}}^{-\frac{1}{\tau }} }{2}-\frac{{\mathrm{e}}^{-\frac{1}{\tau }} -\frac{E_1\left(1/\tau \right)}{\tau }}{2\,\tau }}{3\,\tau }\right)}}{6} \\\tag{\stepcounter{equation}\theequation}
  + (b \,\tau)^4 \, \frac{{\left(\frac{{\mathrm{e}}^{-\frac{1}{\tau }} }{4}-\frac{\frac{{\mathrm{e}}^{-\frac{1}{\tau }} }{3}-\frac{\frac{{\mathrm{e}}^{-\frac{1}{\tau }} }{2}-\frac{{\mathrm{e}}^{-\frac{1}{\tau }} -\frac{E_1\left(1/\tau\right)}{\tau }}{2\,\tau }}{3\,\tau }}{4\,\tau }\right)}}{24} 
- ...\\ 
\end{align*}
\end{small}\noindent Expanding the previous expression and collecting the terms separately involving $E_1(1/\tau)$ and $e^{-\frac{1}{\tau}}$ now gives:
\begin{small}
\begin{equation} \label{eq:14}
\begin{aligned}[b]
W(\tau,b)=& E_1(1/\tau){\left(1+b+\frac{b^2 }{4}+\frac{b^3 }{36}+\frac{b^4 }{576}+...\right)} \\
& +{\mathrm{e}}^{-\frac{1}{\tau }} \left(-b\,\tau -\frac{b^2 \,\tau }{4}-\frac{b^3 \,\tau }{36}-\frac{b^4 \,\tau }{576}+\frac{b^2 \,\tau^2 }{4}\right . \\
& \left . +\frac{b^3 \,\tau^2 }{36}+\frac{b^4 \,\tau^2 }{576}-\frac{b^3 \,\tau^3 }{18}-\frac{b^4 \,\tau^3 }{288}+\frac{b^4 \,\tau^4 }{96}-...\right)
\end{aligned}
\end{equation}
\end{small}
The polynomial that multiplies $ E_1 $  in Eq. \ref{eq:14} can be expressed easily with a power series
\begin{equation}
W_1=E_1\left (1/\tau  \right) \sum \limits_{n=0}^{\infty} \frac{b^n}{n!^2} 
\end{equation}
\noindent For the special case $k=0$, the modified bessel function of the first kind $I_k(z)$  comes down to the power series   
\begin{equation}\label{eq:16}
 I_0(z)=\sum \limits_{n=0}^{\infty}\frac{\left(\frac{z}{2}\right)^{2n}}{n!^2}   
\end{equation}
Thus, $W_1$ can be simplified further by setting $z=2\sqrt{b}$, which gives
\begin{equation}\label{eq:17}
    W_1=E_1\left (1/\tau  \right) I_0\left ( 2\sqrt{b} \right )
\end{equation}

\noindent The second polynomial in Eq. \ref{eq:14} is an alternating power series that can be expressed by a double summation. The second term then simplifies to: 
\begin{equation}\label{eq:18}
 W_2={\mathrm{e}}^{-\frac{1}{\tau }} \,{\left(\sum_{m=0}^{\infty } {{\left(-\tau \right)}}^{m+1} \,m!\,{\left(\sum_{n=m+1}^{\infty } \,\frac{b^n }{{n!}^2 }\right)}\right)}
 \end{equation}
 \noindent It is worthy noting that the summation will converge to a negative value. $W_2$ will however tends towards zero for small evaluation times $\tau$ since $e^{-1/\tau} \rightarrow 0$. On the contrary, for long evaluation times, $e^{-1/\tau}\rightarrow 1$ and $W_2$ will take a negative value.  

Finally, a first analytical expression for the MILS model involving special functions $E_1$ and $I_0$, as well as a double infinite summation is thus provided by:
\begin{small}\begin{equation}\label{eq:19}
\begin{aligned}[b]
\Delta \overline{T}  =  \frac{q\,I_0 \left (2\sqrt{b} \right )}{4 \pi k  }  &\Bigg(  E_1\left( \frac{1}{\tau }\right)\,{{\textrm{I}}}_0 \left(2\,\sqrt{b}\right) \\
 & + {\mathrm{e}}^{-\frac{1}{\tau }} \,{\sum_{m=0}^{\infty } \sum_{n=m+1}^{\infty } {{\left(-\tau \right)}}^{m+1} \,m!\,{ \frac{b^n }{{n!}^2 }}} \Bigg)
\end{aligned}
\end{equation}\end{small}

\begin{figure*}
 \centering
    \includegraphics[width=\textwidth,trim={2.25cm 1cm 2.cm 1.25cm},clip]{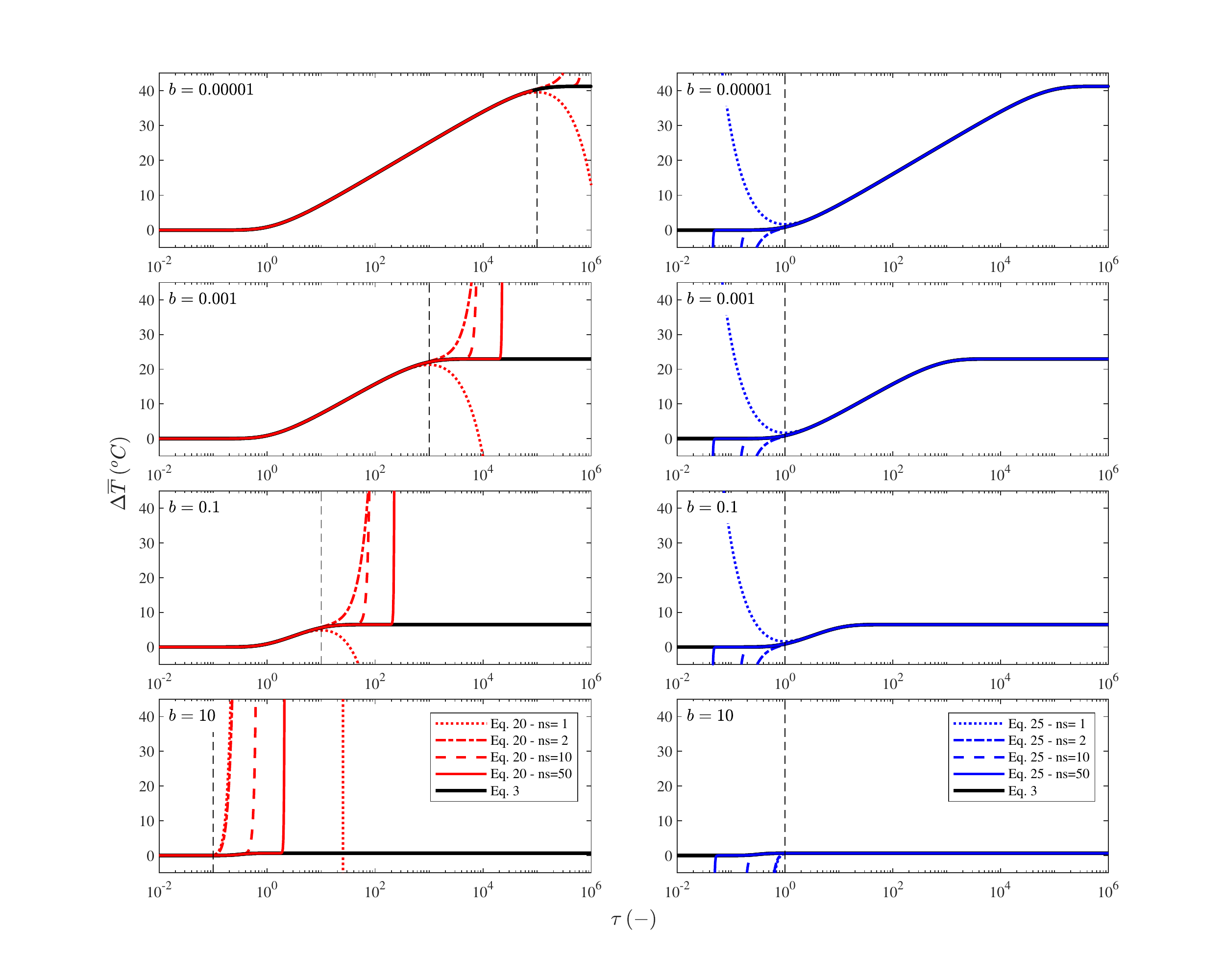}
 \caption{Mean temperature change $\Delta\overline{T}(b,\tau)$ obtained by Eqs. \ref{eq:3}, \ref{eq:19} (left) and \ref{eq:24} (right) for $k=2$ W/mK, $q$=100 W/m and various values of $b$ and number of summands $ns$. The vertical black dashed lines correspond to $\tau=1/b$ (left) or $\tau=1$ (right). \textcolor{black}{Dimensionless parameter $b=(P\acute{e}/4)^2$ is proportional to the ratio of advective to conductive heat transfer, while $\tau=4Fo$ is proportional to the ratio of conductive heat transfer to the rate of heat storage. }}
 \label{FIG:2}
\end{figure*}

\begin{figure}
 \centering
    \includegraphics[height=6.5cm]{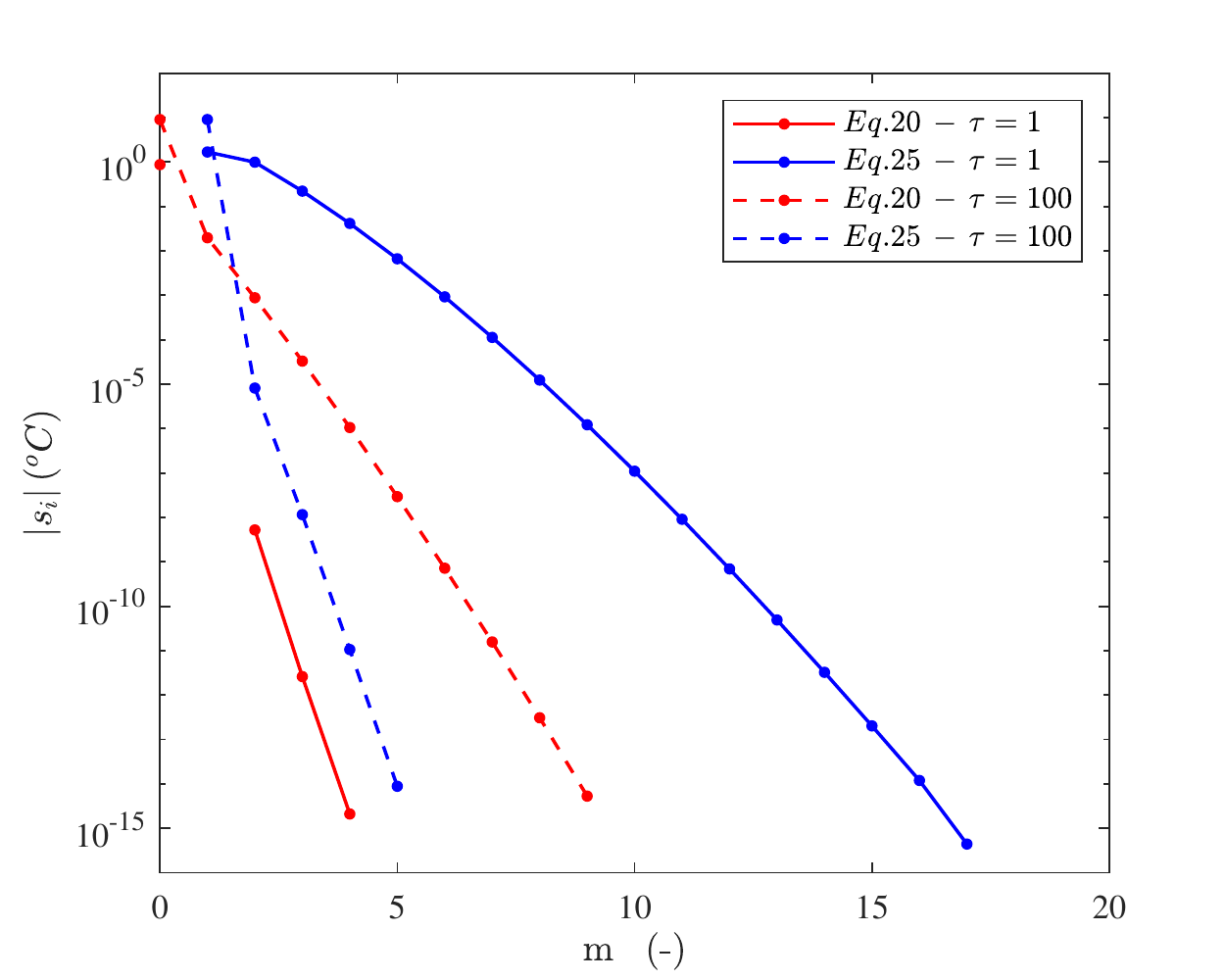}
 \caption{Absolute value of the summands $s_i$ \textcolor{black}{obtained with eqs. \ref{eq:19} and \ref{eq:24}} as a function of summation index $m$ for $b=0.001$, $\tau=1$ and  $\tau=100$.}
 \label{FIG:4}
\end{figure}

\noindent It is interesting to note that without groundwater flow, $b=0$, $I_0(0)=1$ and Eq. \ref{eq:19} corresponds exactly to the infinite line source model of \cite{ingersoll_heat_1954}. For $b>0$, $I_0(2b^{1/2})>1$ and the temperature plateau typically observed at the long-term (see Fig. \ref{FIG:2}) will be controlled mostly by the negative values taken by $W_2$.  We will show in Section \ref{Sec:4} that this equation is appropriate for $0< \tau \leq 1/b $. 
For this range, one can show quite easily that the absolute value of the inner sum decreases monotonically as $m$ increases and tends towards zero for $m \rightarrow \infty$.  To this end, see in Fig. \ref{FIG:4} how rapidly the summands $s_i$ tend to machine precision.  Thus, the alternating series test confirms convergence of the power series for $0< \tau \leq 1/b $.

\subsection{Analytical expression for $\boldsymbol{\tau \geq 1}$ }

A second expression is obtained by developing the summation in Eq. \ref{eq:9} and reducing the order of the generalized exponential integral until only $E_1$ remains. Finally, collecting separately the terms involving $E_1(b\tau)$ and $e^{-b\tau}$ leads to:
\begin{small}
\begin{equation} \label{eq:20}
\begin{aligned}[b]
\sum _{n=0}^{\infty} & \frac{1  }{(-\tau)^n n!}  E_{n+1}(b\tau) = \\
& E_1(b\tau) {\left(1+b+\frac{b^2 }{4}+\frac{b^3 }{36}+\frac{b^4 }{576}+...\right)} \\ 
& \quad +e^{-b\tau}\Bigg(
-\frac{1+\frac{b}{4}+\frac{b^2 }{36}+\frac{b^3 }{576}+...}{\tau } \\
& \quad\quad\quad\quad\quad +\frac{\frac{1}{4}+\frac{b}{36}+\frac{b^2 }{576}+...}{\tau^2 } \\
& \quad\quad\quad\quad\quad\quad -\frac{\frac{1}{18}+\frac{b}{288}+...}{\tau^3 }\\
& \quad\quad\quad\quad\quad\quad +\frac{\frac{1}{96}+...}{\tau^4 }+... 
\Bigg)
\end{aligned}
\end{equation}
\end{small}Using again the identity of Eq. \ref{eq:16} allows simplifying the first polynomial of Eq. \ref{eq:20} to
\begin{equation} \label{eq:21}
\begin{aligned}[b]
W_1=E_1\left (b \tau  \right) I_0\left( 2\sqrt{b} \right )
\end{aligned}
\end{equation}
The second polynomial can be expressed by a double summation and the resulting expression is 
\begin{equation} \label{eq:22}
\begin{aligned}[b]
W_2={\mathrm{e}}^{-b\tau} \,{\sum_{m=1}^{\infty } \sum_{n=0}^{\infty } {{\frac{b^n(m-1)!}{(-\tau)^{m}(m+n)!^2}} }}
\end{aligned}
\end{equation}
which can be simplified further using the generalized hypergeometric function $_1F_2$:
\begin{equation} \label{eq:23}
\begin{aligned}[b]
W_2=e^{-b\tau}\sum_{m=1}^{\infty } \frac{_1 {{\textrm{F}}}_2 \left(1;\;m+1,m+1;\;b\right)}{{{\left(-\tau \right)}}^m\, m \,m!}
\end{aligned}
\end{equation}

Numerical tests indicated that Eq. \ref{eq:22} was 1 to 2 orders of magnitude faster than Eq. \ref{eq:23}. For performance purposes, a second expression for $\overline{T}$ is thus provided by:
\begin{small}\begin{equation} \label{eq:24}
\begin{split}
\Delta \overline{T}  =\frac{q\,I_0\left( 2\sqrt{b}\right)}{4\pi k  } &\Bigg( 2K_0 \left(2\sqrt{b} \right )
 -  E_1\left(b\tau\right)\,{{\textrm{I}}}_0 \left(2\,\sqrt{b}\right) \\
& - {\mathrm{e}}^{-b\tau} \,{\sum_{m=1}^{\infty } \sum_{n=0}^{\infty } {{\frac{b^n(m-1)!}{(-\tau)^{m}(m+n)!^2}} }}
\Bigg)
\end{split}
\end{equation}\end{small}

\noindent As $E_1(0)=\infty$, it is this time impossible to identify the infinite line source model for the special case $b=0$. The physical interpretation of Eq. \ref{eq:24} is however instructive.  For long evaluation times $\tau\rightarrow \infty$, $W_1$ and $W_2$ will be nil and the steady-state temperature plateau will be equal to $\overline{T}=q I_0(2b^{1/2})K_0(2b^{1/2})/2\pi k$, a result already obtained by \cite{sutton_ground_2003} and \cite{diao_heat_2004}. Since $_1F_2(1;m+1,m+1;b)\rightarrow 1$ as $m \rightarrow \infty$, the alternating series test indicates again convergence of the power series for $\tau \geq 1 $ and for any $b$ value. We will show in Section \ref{Sec:4} that Eq. \ref{eq:24} is appropriate for $\tau \geq 1 $. 

\subsection{First-order approximations}\label{Sec:}

First-order approximations are often useful for practical applications. In particular, they allow analysis of thermal response tests by simplifying the temperature \citep{mogensen_fluid_1983} or time derivative response \citep{pasquier_interpretation_2018}. Such approximations are obtained for the MILS model using a single summand in the power series and the relations $I_0(x) \approx 1+x$, $K_0(x) \approx -2\ln{(x)}$ and $E_1(x) \approx -\ln{(x)}-\gamma$. Using these relations, one can derive the following first-order approximation easily from Eqs. \ref{eq:17} and \ref{eq:18}:
\begin{equation}\label{eq:19FOA}
\tilde{W}   =\left(\ln{(\tau)}-\gamma \right)(1+b) - b\tau e^{-\frac{1}{\tau}}
\end{equation}
\noindent Similarly, a second approximation is derived from Eqs. \ref{eq:21} and \ref{eq:22}:
 \begin{equation}\label{eq:24FOA}
\tilde{W}  =   \left(\ln{\left( b\tau\right)}+\gamma\right)(1+b)+ \frac{1}{\tau}e^{-b\tau}  - 2\ln{\left(2\sqrt{b}\right)} 
\end{equation}
\noindent Note that the same formulas would have been obtained using Eqs. \ref{eq:8} and \ref{eq:9} with two summands, and by expressing $E_2$ as a function of $E_1$ with the recursion formula presented earlier. The accuracy of these approximations will be presented in the next Section. 

\section{Numerical evaluation and efficiency}\label{Sec:4}
\begin{figure*}
 \centering
    \includegraphics[width=0.975\textwidth,trim={2.cm 1.05cm 2.cm 1.45cm},clip]{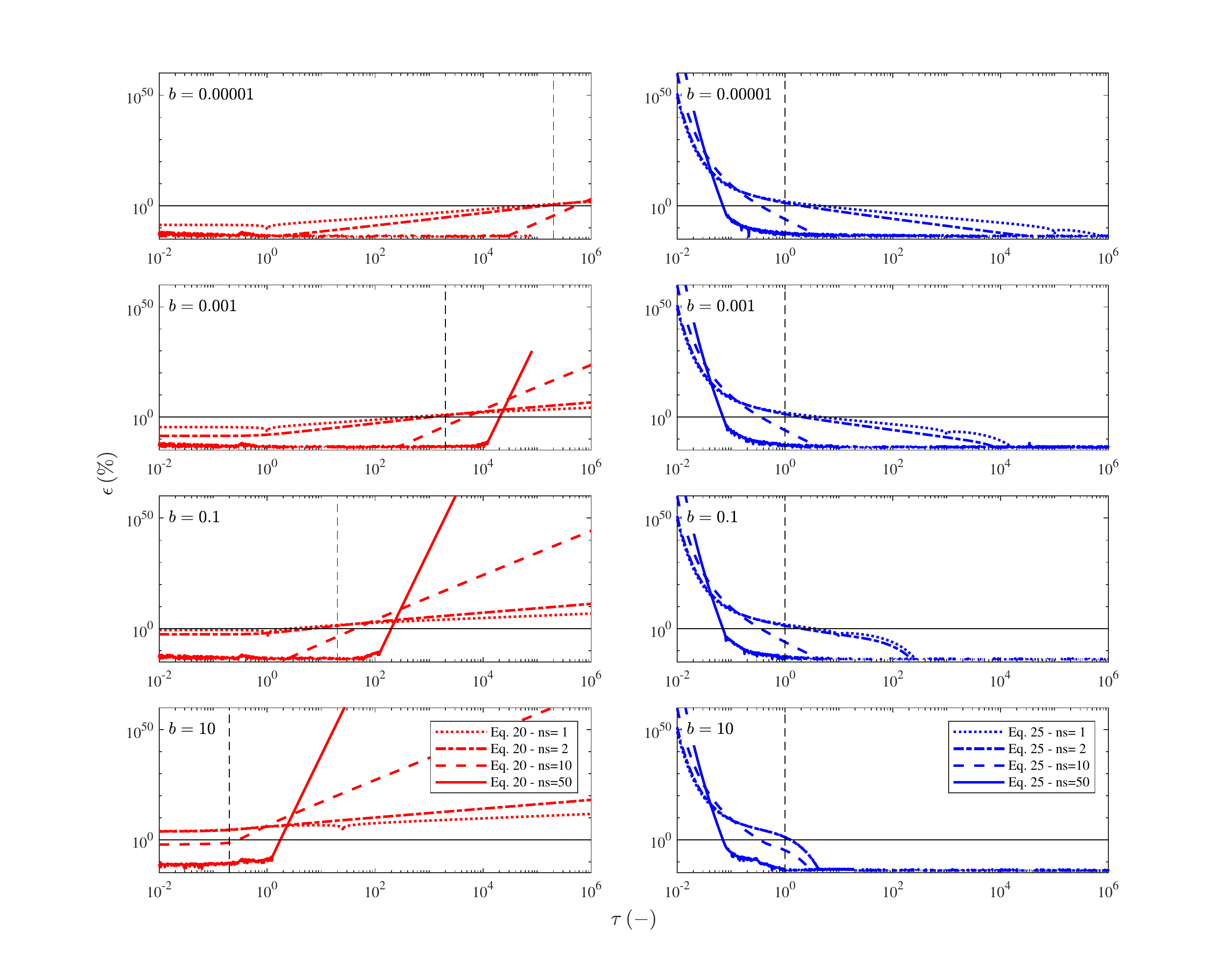}

 \caption{Relative absolute error $\epsilon$ made on $W$ when using Eqs. \ref{eq:19} (left) and \ref{eq:24} (right) with respect to the integral of Eq. \ref{eq:3}, for various values of $b$ and number of summands $ns$. The vertical black dashed lines correspond  to $\tau=1/b$ (left) or $\tau=1$ (right). The thin horizontal black lines correspond to an error of 1\%.  \textcolor{black}{Dimensionless parameter $b=(P\acute{e}/4)^2$ is proportional to the ratio of advective to conductive heat transfer, while $\tau=4Fo$ is proportional to the ratio of conductive heat transfer to the rate of heat storage.  }}
 \label{FIG:3}
\end{figure*}

The mean temperature change $\Delta \overline{T}(b,\tau)$ obtained by Eqs. \ref{eq:19} and \ref{eq:24} for a finite number of summands are illustrated in Fig. \ref{FIG:2} for various values of $b$ and for $10^{-2} \leq \tau \leq 10^6$. This range covers a few minutes to several years for radial distances $r$ and thermal diffusivities $\alpha$ typical of geothermal applications. Both the inner and outer summations were evaluated with the same number of terms. In addition, the relative error $\epsilon$ made by Eqs. \ref{eq:19} and \ref{eq:24} with respect to quadrature of Eq. \ref{eq:3} is shown in Fig. \ref{FIG:3} for partial sums made of 1, 2, 10 and 50 summands. Note that both figures contain a vertical line at $\tau=1/b$ or $\tau=1$ that represents the convergence criteria of the power series mentioned previously. These values were chosen to ensure an error less than 1\% when only 10 summands are used to evaluate the infinite power series. \textcolor{black}{ Recall that $\tau=4Fo$ and $b=(P\acute{e}/4)^2$. The results presented hereinafter can therefore also be expressed as a function of the dimensionless numbers $Fo$ and $P\acute{e}$.  }

One can see easily in Fig. \ref{FIG:2} and \ref{FIG:3} that both analytical expressions converge rapidly with an error inferior to 1\% when only 10 terms are used to approximate the infinite power series. As illustrated in Fig. \ref{FIG:4}, the summands decrease and reach machine precision rapidly. A summand is then evaluated numerically equal to zero and adding the subsequent terms of the series is then useless. This confirms that in practice just a few terms are required to obtain a good numerical accuracy. However, this result is only valid if the convergence criteria are met. Otherwise, the series  increase rapidly and diverge with relative errors that could easily exceed $10^{10}\%$. For the special case $b=0$, Eq. \ref{eq:24} will return an infinite value due to the evaluation of $E_1(0)$ and $K_0(0)$.  Formula \ref{eq:19}, which simplifies to the infinite line source model for $b=0$, should then be used. We mention that we observed similar convergence properties and relative errors with Hunt's original solutions (Eqs. \ref{eq:8} and \ref{eq:9}). One can then use both approaches. 
% trim={<left> <lower> <right> <upper>}
It is interesting to note that the series are valid for $\tau\leq 1/b$ or $\tau \geq 1$ and overlap over a significant validity range \textcolor{black}{for most values of $b$}. In practice, one can then use the convergence criteria to create two smaller $\tau$ sub-vectors and thus accelerate the evaluation of equations \ref{eq:19} and \ref{eq:24}. Combination of the individual solutions allows easily to cover the full simulation span. A Matlab implementation of Eqs. \ref{eq:19} and \ref{eq:24} has been developed (see Appendix). The mean computation time for a vector $\tau$ composed of 10 000 time steps with $n_s=10$ was approximately 13 ms, which is 200 times faster than the numerical quadrature of the Hantush well function. The implementation Eqs. \ref{eq:19} and \ref{eq:24} was 15 to 20 \% faster than Hunt's solutions, a relatively small but significant real gain.

\begin{figure}[h]\label{FIG:5}
	\centering
    \includegraphics[width=0.5\textwidth,trim={2.25cm 10cm 12.5cm 1.25cm},clip]{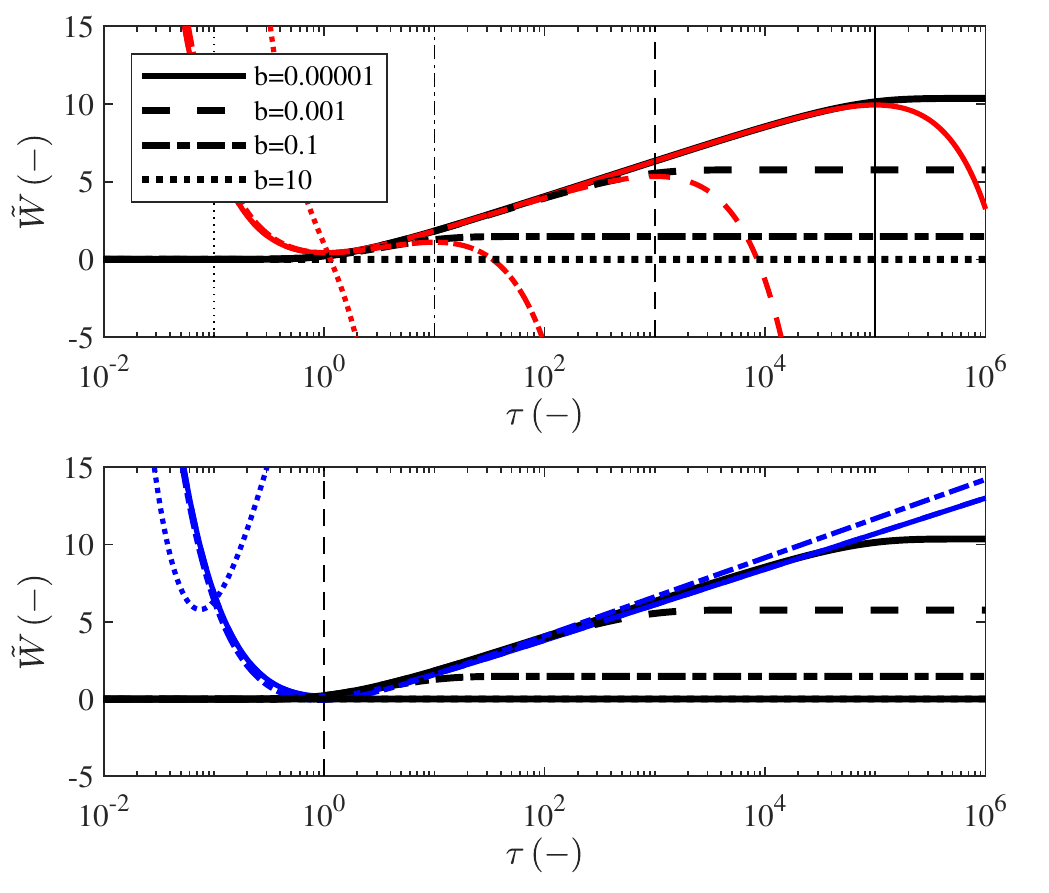}
	\caption{First-order approximations $\tilde{W}$ provided by Eqs. \ref{eq:19FOA} (up) and \ref{eq:24FOA} (down) with the reference solution provided by Eq. \ref{eq:3}. The vertical black lines correspond to $\tau=1/b$ (up) or $\tau=1$ (down). \textcolor{black}{The line $\tau=1/b$ divides the conduction-dominated and advection-dominated regions.}} 
\end{figure}

\begin{figure}[h]
	\centering
    \includegraphics[height=7.5cm,trim=30 0 0 0, clip]{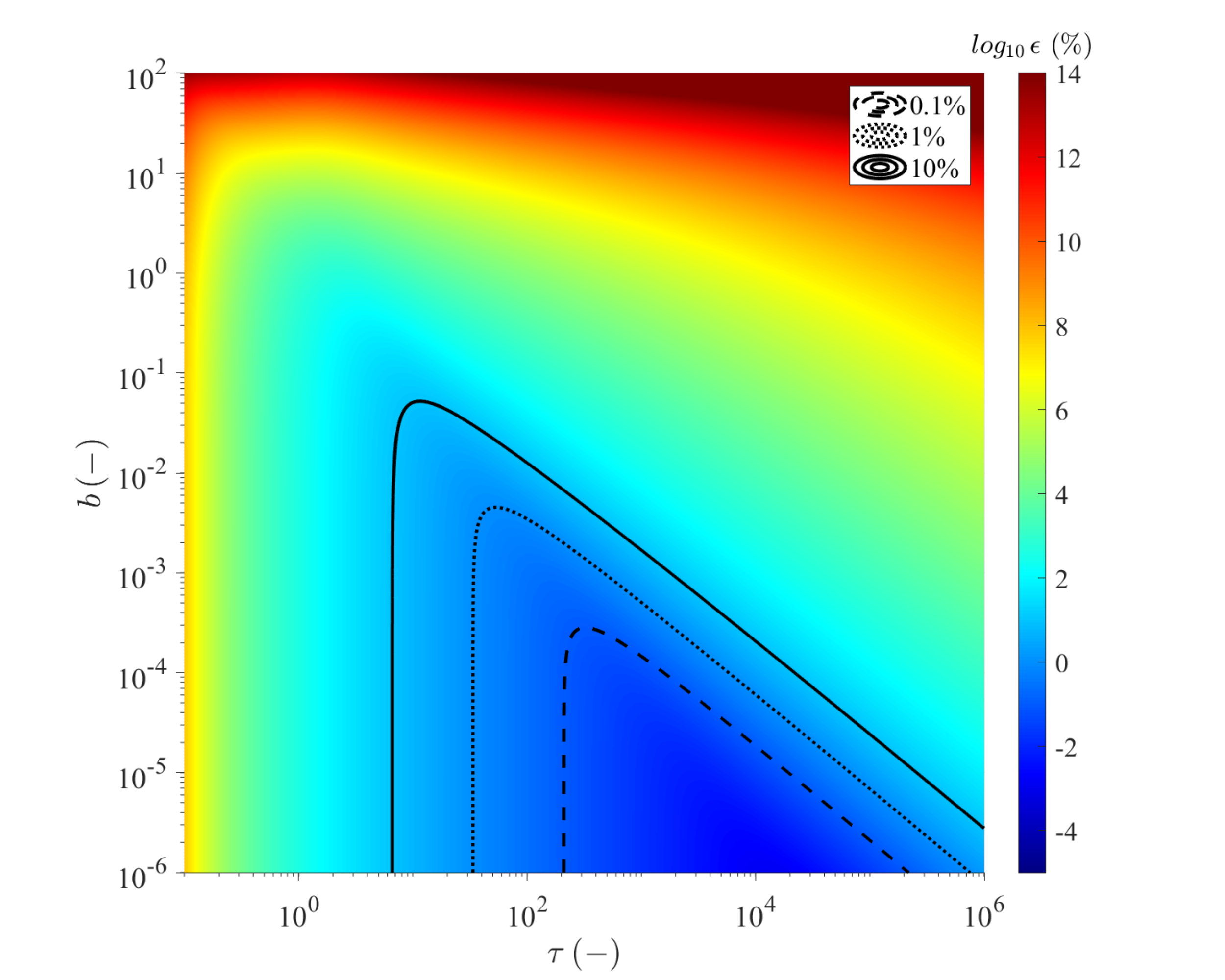}
	\centering
    \includegraphics[height=7.5cm,trim=30 0 0 0, clip]{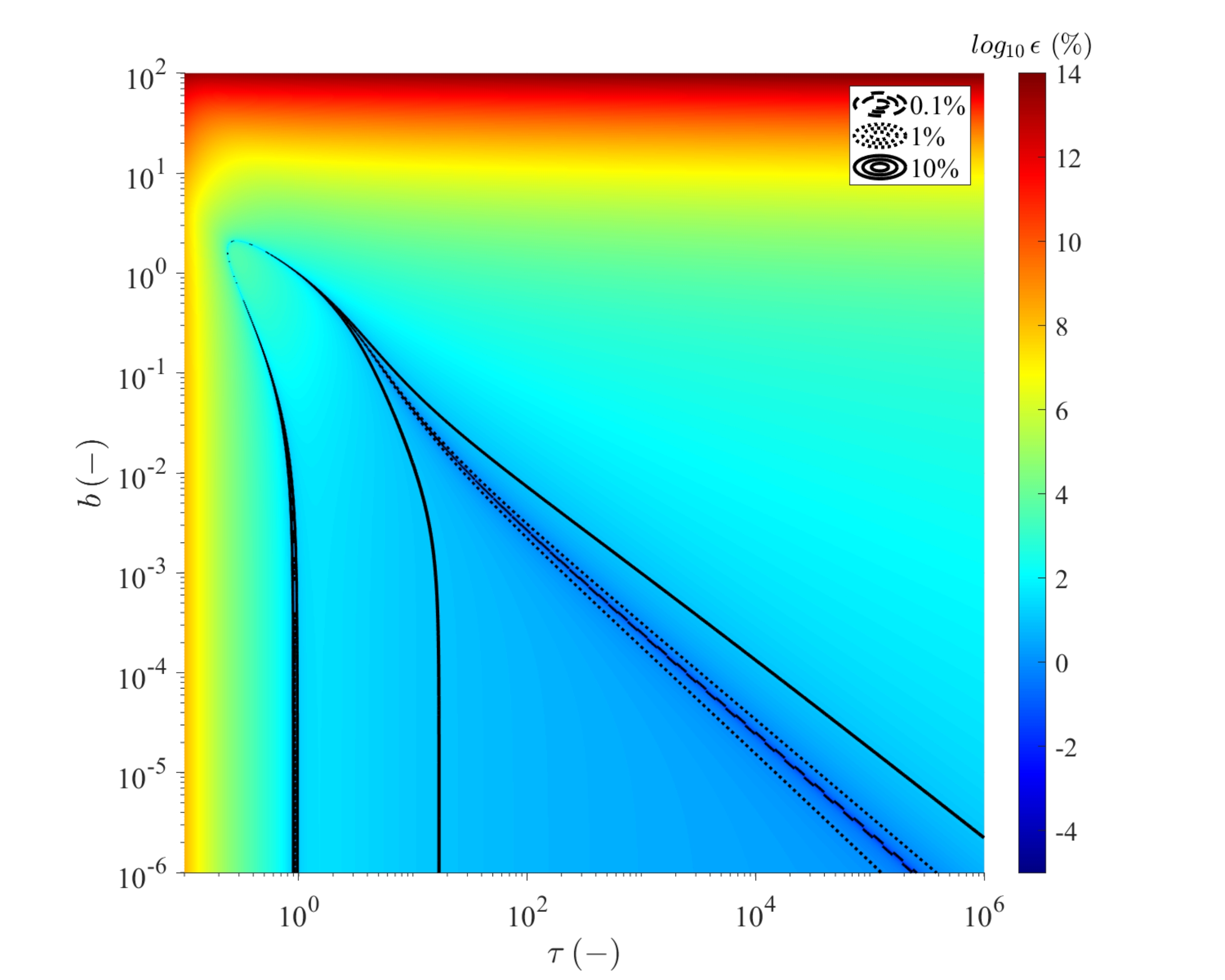}
	\caption{Absolute relative error $\epsilon$ made by the first-order approximations $\tilde{W}$ of Eqs. \ref{eq:19FOA} (up) and \ref{eq:24FOA} (down) with respect ot the reference solution provided by Eq. \ref{eq:3}. The contour lines correspond to error of 0.1\%, 1\% and 10\%.}
	\label{FIG:6}
\end{figure}

Finally, a comparison between the first-order approximation of the Hantush well function and its reference value is illustrated in Fig. \ref{FIG:5} for several values of $b$. If we except the case $b=10$, the approximations provided by Eqs. \ref{eq:19FOA} and \ref{eq:24FOA} appear valid during the conduction-dominant period, \textcolor{black}{ that is during the rising part of $W$ and before the plateau is reached.} For early times and advective-dominant periods, the approximations clearly failed to reproduce $W$. As shown in Fig. \ref{FIG:6}, a relative error of less than 10\% is however achievable and covers two wide isosceles triangles. A vertical narrow valley of small errors around $\tau=1$ is also observed for Eq. \ref{eq:24FOA}. To guide the use of the approximations, the triangular areas are limited for error levels of 0.1\%, 1\% and 10\% by $\tau \geq\tau_{min} \cap \tau \leq \tau_{max}$, where $\log_{10}(\tau_{min})$ and $\log_{10}(\tau_{max})$ are given by the parameters listed in Table \ref{tab:1}. Using these simple relations directly provides the maximum error made by the first-order approximations, which can be useful for practical or pedagogical applications.
\begin{table}[]
\centering
\caption{Numerical values for $a_0$, $a_1$ and $a_2$ used with $\log_{10}(\tau_{min} \, or \,\tau_{max})=(a_0+(\log_{10}(b)-a_1)/a_2)$.}
\label{tab:1}
\begin{scriptsize}
\begin{tabular}{c c c c c c } 
\hline \hline 
\multirow{2}{2em}{ } & \multirow{2}{2em}{$\epsilon$}  & \multirow{2}{2em}{Bound} & \multirow{2}{2em}{$a_0$ } & \multirow{2}{2em}{$a_1$} & \multirow{2}{2em}{$a_2$} \\
& & & & & \\
\hline 
\multirow{6}{3em}{Eq. \ref{eq:19FOA}} & \multirow{2}{2em}{0.1\%} & $\tau_{min}$ & 2.32 & 0 & $\infty$  \\ 
&  & $\tau_{max}$ & 0 & -1.04 & -0.93  \\ 
& \multirow{2}{2em}{1\%} & $\tau_{min}$ & 1.52 & 0 & $\infty$  \\ 
&   & $\tau_{max}$ & 0 & -0.46 & -0.94  \\ 
& \multirow{2}{2em}{10\%} & $\tau_{min}$ & 0.81 & 0 & $\infty$  \\ 
&   & $\tau_{max}$ & 0 & 0.08 & -0.94  \\ 
\hline
\multirow{4}{3em}{Eq. \ref{eq:24FOA}} & \multirow{2}{2em}{1\%} & $\tau_{min}$ & 0 & -0.53 & -1.07  \\ 
&   & $\tau_{max}$ & 0 & -0.55 & -0.97  \\ 
& \multirow{2}{2em}{10\%}  & $\tau_{min}$ & 1.22 & 0 & $\infty$  \\ 
&   & $\tau_{max}$ & 0 & -0.36 & -0.88  \\ 
\hline \hline\end{tabular} \end{scriptsize}
\end{table}

\section{Conclusions}
We presented analytical expressions and approximations for the moving infinite line source model, a widely used thermal model that takes into account groundwater flow. All expressions were derived from the Hantush well function and generalized incomplete gamma function that are commonly used in hydrogeology to model leaky confined aquifers. 

The proposed new analytical expressions are exact, integral-free and take the form of convergent power series involving no recursive evaluations. We have shown that for most cases, providing the convergence domain is respected, evaluation of the series with only a few summands provides efficiently a high accuracy.

New first-order approximations were also presented along with their validity domain. It was shown that a relative error less than 10\% can be obtained easily with these approximations which can be useful for many professional or academic activities.

\textcolor{black}{To conclude, we stress that the approach used in this work is not limited to the moving infinite line source model and could probably lead to new analytical expressions for other models that currently require a heavy numerical quadrature. Future work could then lead to new series expansion useful to the design of ground heat exchangers.}

\section*{Acknowledgments}
The authors wish to thank the anonymous referees who reviewed this work and provided helpful and constructive comments and suggestions.  This work was financed by the Natural Sciences and Engineering Research Council of Canada through grant number RGPIN-2019-04713.

\appendix
\section*{Matlab Implementation of the MILS model} \label{Appendix}
\begin{scriptsize}
\begin{lstlisting} 
function [dT]=MILS(t,k,Cs,Cw,vD,r,q,ns)

% t : Column vector of evaluation times [s]
% k : Thermal conductivity [W/(mK)]
% Cs: Soil volumetric capacity [J/(m^3 K)]
% Cw: Water volumetric capacity [J/(m^3 K)]
% vD: Darcy velocity/groundwater flux [m/s]
% r : Radial distance [m]
% q : Normalized heat flux [W/m]
% ns: Number of summands [-]
% dT: Mean temperature change [K]. The columns
%     correspond to Eq. 20 and 24 respect.

m=0:ns-1;
n=m+1;

b=(r*vD*Cw/(4*k)).^2;
tau=4*k/Cs*t/r.^2;
I0=besseli(0,2*sqrt(b));
a0=q*I0/(4*pi*k);
id=sparse(tril(toeplitz(n))>0);

% Implementation of Eq. 20
S2=sum(repmat(b.^n./factorial(n).^2,ns,1)'.*id,1);
S1=sum((-tau).^(m+1).*factorial(m).*S2,2);
T(:,1)=a0*(expint(1./tau)*I0+exp(-1./tau).*S1);

% Implementation of Eq. 25
S2=b.^(n-1)'./factorial((m+1)+(n-1)').^2;
S1=sum(factorial(m)./(-tau).^(m+1).*sum(S2,1),2);
T(:,2)=a0*(2*besselk(0,2*sqrt(b))...
    -expint(b*tau)*I0-exp(-b*tau).*S1);

end %end of function
\end{lstlisting}
\end{scriptsize}
% VARIABLES
\nomenclature[V]{$ \alpha $}{ Thermal  diffusivity [ $m^2/s$ ]}
\nomenclature[V]{$ \beta $}{$2 \alpha /v_T$ [ m ]}
\nomenclature[V]{$ \psi $}{ Integration variable [ - ]}
\nomenclature[V]{$ \tau $}{ Adimensional time ($4\alpha t/r^2$) [ - ]}
\nomenclature[V]{$ \theta $}{ Angle with the x-axis [ rad ]}
\nomenclature[V]{$ b $}{$(rv_T/4\alpha)^2$ [ - ]}
\nomenclature[V]{$ C $}{ Ground volumetric heat capacity [ J/(m$^3\cdot$K)]}
\nomenclature[V]{$ C_w $}{ Water volumetric heat capacity [ J/(m$^3\cdot$K)]}
\nomenclature[V]{$ E_n $}{ Generalized exponential integral function [ - ]}
\nomenclature[V]{$ F_o $}{ Fourier number ($\alpha t/r^2$) [ - ]}
\nomenclature[V]{$ I_0 $}{ Modified bessel function of the first kind and of order 0 [ - ]}
\nomenclature[V]{$ k $}{ Ground thermal  conductivity [ W/(m$\cdot$K) ]}
\nomenclature[V]{$ K_0 $}{ Modified bessel function of the second kind and of order 0 [ - ]}
\nomenclature[V]{$ m $}{ Summation index [ - ]}
\nomenclature[V]{$ n $}{ Summation index [ - ]}
\nomenclature[V]{$ ns $}{ number of summands [ - ]}
\nomenclature[V]{$ P\acute{e} $}{ Péclet number ($r v_T/\alpha $) [ - ]}
\nomenclature[V]{$ q $}{ Normalized heat load [ W/m ]}
\nomenclature[V]{$ r $}{ Radial distance [ m ]}
\nomenclature[V]{$ \Delta T $}{ Temperature change [ $^{\circ}$C ]}
\nomenclature[V]{$ \Delta \overline{T} $}{ Mean temperature change[ $^{\circ}$C ]} 
\nomenclature[V]{$ u $}{$r^2/4 \alpha t$ [ - ]}
\nomenclature[V]{$ v_D $}{ Darcy velocity or groundwater flux [ m/s ]}
\nomenclature[V]{$ v_T $}{ Effective heat transport velocity [ m/s ]}
\nomenclature[V]{$ W $}{ Hantush well function [ - ]}

% ACRONYMS
\nomenclature[A]{$ GHE $}{ Ground heat exchanger}
\nomenclature[A]{$ MILS $}{Moving infinite line source}
\printnomenclature

\printcredits

\bibliographystyle{cas-model2-names}

% \bibliography{ReferencesPasquierSep2021}
% \bibliography{Biblio_PPasquier_Novembre_2021b}
\bibliography{Biblio_PPasquier_Septembre_2021a}

 
\end{document}